\documentclass[12pt,preprint,notoc,nohyper]{AKRSX} 

\usepackage{epsfig,multicol,bbm}

\newcommand\fverb{\setbox\pippobox=\hbox\bgroup\verb}
\newcommand\fverbdo{\egroup\medskip\noindent%
            \fbox{\unhbox\pippobox}\ }
\newcommand\fverbit{\egroup\item[\fbox{\unhbox\pippobox}]}
\newbox\pippobox

\title{Curvature and Weyl collineations of spacetimes}

\author{A. R. Kashif$^1$\footnote{\emph{Present address:} Department of Mathematics (Prep year),
P. O. Box 2440, University of Hail, Saudi Arabia.}~, K.
Saifullah$^2$ \\

$^1$College of EME, National University of Sciences and Technology,
Rawalpindi, Pakistan\\

$^2$Department of Mathematics, Quaid-i-Azam University, Islamabad,
Pakistan\\

Electronic address: \email{kashmology@yahoo.com},
\email{saifullah@qau.edu.pk}}

\preprint{}  

\abstract{Lie symmetries of various geometrical and physical
quantities in general relativity play an important role in
understanding the curvature structure of manifolds. The Riemann
curvature and Weyl tensors are two fourth-rank tensors in the
theory. Interrelations between the symmetries of these two tensors
(known as collineations) are studied. Some illustrative examples are
also provided.}

\dedicated{}

\begin{document}

Let us consider a smooth four-dimensional manifold $M$ on which $g$
is a spacetime metric.  On this manifold $R_{bcd}^{a}$ represents
the Riemann curvature tensor, and its trace $R_{ab}$ is the Ricci
tensor. The Lie symmetries of these tensors play an important role
in understanding the curvature structure of spacetimes in general
relativity \cite{9r, 2r}.  Symmetries of the metric tensor, called
\textit{isometries} or \textit{Killing vectors} (KVs), are given by
\begin{equation}\label{1}
\pounds _{_{\mathbf{X}}}\mathbf{g}=0,
\end{equation}
where $\pounds _{_{\mathbf{X}}}$ is the Lie derivative along the
vector field $\mathbf{X}$. Putting $\phi \mathbf{g}$ on the right
side of Eq. (\ref{1}), where $\phi$ is a function of the
coordinates, gives the \textit{conformal Killing vectors}. However,
if $\phi$ is a constant $\mathbf{X}$ is called a \textit{homothetic
vector} (HV). Replacing $\mathbf{g}$ by any other tensor field gives
the symmetries of that tensor, called \textit{collineations}
\cite{kit}. Thus curvature collineations (CCs) are the symmetries of
$R_{bcd}^{a}$, and Weyl collineations (WCs) are those of the Weyl
tensor

\[
C_{cd}^{ab}=R_{cd}^{ab}-2\delta _{\lbrack
c}^{[a}R_{d]}^{b]}+\frac{1}{3}\delta _{\lbrack c}^{a}\delta
_{d]}^{b}R .
\]

The Weyl tensor describes the purely gravitational field for a given
space. Mathematically the curvature and the Weyl tensors have
similar forms, but their symmetries (i.e. CCs and WCs) are
different. If we replace $\mathbf{g}$ by the curvature tensor, Eq.
(1) defines CCs. This can be written in component form as
\[
R_{bcd,f}^{a}X^{f}+R_{fcd}^{a}X_{,b}^{f}+R_{bfd}^{a}X_{,c}^{f}+
R_{bcf}^{a}X_{,d}^{f}-R_{bcd}^{f}X_{,f}^{a}=0,
\]
where ``$,$'' denotes the partial derivative. Similarly we define
WCs. While the metric tensor is always non-singular and KVs form a
finite dimensional Lie algebra, CCs and WCs can give rise to finite
as well as infinite dimensional Lie algebras. The curvature tensor
can be represented by a $6\times6$ matrix on account of its
algebraic symmetries, whose rank gives the rank of the tensor.
Similarly the Weyl tensor can also be written as a $6 \times 6$
matrix. It is possible that its rank is greater or less than the
rank of the corresponding curvature matrix. If the rank of the
curvature matrix is $\geq 4$ then the Lie algebra of CCs is finite
dimensional \cite{Ahb-Ark-Aq}. Clearly the set of KVs is contained
in the set of WCs and CCs, but the question arises about the
relationship between WCs and CCs.

An example of a spacetime is provided in Ref. \cite{ibrar} for which
WCs are properly contained in the CCs. Consider the plane symmetric
metric
\begin{equation}\label{3}
ds^{2}=e^{\nu (x)}dt^{2}-dx^{2}-e^{\mu (x)}(dy^{2}+dz^{2}) .
\end{equation}
It has 4 KVs

\[
\mathbf{X}_{0}=\frac{\partial }{\partial t}, \mathbf{X}_{1}=\frac{%
\partial }{\partial y}, \mathbf{X}_{2}=\frac{\partial }{\partial z},
\mathbf{X}_{3}=-z\frac{\partial }{\partial y}+y\frac{\partial
}{\partial z}.
\]

When $\nu =0$ and $e^{\mu }=(r/a)^{2}$, the non-zero component of
the curvature tensor is
\[
R_{323}^{2}=-\frac{1}{a^{2}} ,
\]
and those of the Weyl tensor are
\begin{eqnarray*}
C_{101}^{0} &=&-\frac{1}{3r^{2}},
C_{202}^{0}=\frac{1}{6}=C_{212}^{1},
\\
C_{303}^{0} &=&\frac{1}{6a^{2}}=C_{313}^{1},
C_{323}^{2}=-\frac{1}{3a^{2}}.
\end{eqnarray*}
Here the Ricci tensor is of rank 2 and is degenerate. This case has
one extra KV
\[\mathbf{X}_{3}=-\frac{z}{a}\frac{\partial}{\partial \theta}+a\theta \frac{\partial }{\partial z},\]
one proper HV
\[
\mathbf{X}_{4}=t\frac{\partial }{\partial t}+r\frac{\partial
}{\partial r},
\]
and one additional WC
\[
\mathbf{X}_{5}=\frac{1}{2}(t^{2}+r^{2})\frac{\partial }{\partial t}+tr\frac{%
\partial }{\partial r}.
\]
Thus CCs form an infinite dimensional Lie algebra, and therefore WCs
are properly contained in the CCs.

Now we take another example, with $e^{\nu }=e^{\mu }=(x/a)^{b}$
($a,b\neq 0\in \Bbb{R}$) in Eq. (\ref{3}). The non-zero components
of the curvature tensor are
\begin{eqnarray*}
R_{101}^{0} &=&\frac{b}{4}(\frac{2-b}{x^{2}}),
R_{212}^{1}=R_{313}^{1}=(\frac{x}{a})^{b}(\frac{b(2-b)}{4x^{2}}), \\
R_{202}^{0}
&=&\frac{-b^{2}x^{b}}{4a^{b}x^{2}}=R_{303}^{0}=R_{323}^{2}.
\end{eqnarray*}
It has two additional KVs
\[
\mathbf{X}_{4}=z\frac{\partial }{\partial t}+t\frac{\partial
}{\partial z},
\mathbf{X}_{5}=y\frac{\partial }{\partial t}+t\frac{\partial }{%
\partial y},
\]
and one proper HV which is also a CC
\[
\mathbf{X}_{6}=(t\frac{\partial }{\partial t}+y\frac{\partial }{\partial y}+z%
\frac{\partial }{\partial z}), (b\neq 2).
\]
It has infinitely many WCs and every vector field is a WC. The CCs
are also infinite dimensional but not every vector field is a CC.
Thus CCs are contained in WCs.

Some interesting cases of the interrelations between CCs and WCs
given in Ref. \cite{HallKashif} are worth mentioning here:

\begin{enumerate}
\item[a.] Spacetimes exist for which CCs are properly contained in WCs, with either CCs and WCs
each finite dimensional, each infinite dimensional or with CC finite
dimensional and WC infinite dimensional. Examples also exist when
both are equal in the finite- as well as infinite-dimensional cases.

\item[b.] Spacetimes exist for which WCs are properly contained in CCs, with either WCs and
CCs each finite dimensional, each infinite dimensional or with WCs
finite dimensional and CCs infinite dimensional.

\end{enumerate}

An interesting example of a case is also provided in Ref.
\cite{HallKashif} where neither of the sets of CCs and WCs is
contained in the other.

\acknowledgments

ARK and KS acknowledge travel grants, respectively from the National
University of Sciences and Technology, Islamabad, and the Higher
Education Commission of Pakistan, to participate in MG12 in July
2009, at UNESCO, Paris.

\end{document}